\documentclass[aps,12pt,pra,showpacs,superscriptaddress,preprint]{revtex4}

\usepackage{amsmath}

\usepackage[dvipdfmx]{graphicx}

\usepackage{subfig}

\usepackage{array}

\usepackage{amsfonts}

\usepackage{epsf}

\usepackage{epsfig}

\usepackage{amssymb}

\usepackage{bbm}

\usepackage{delarray}

\usepackage{caption}

\usepackage{amstext}

\usepackage{graphics}

\usepackage{epstopdf}

\catcode`ð=\active

\defð{\u{g}}

\catcode`Ð=\active

\defÐ{\u{G}}

\catcode`Ý=\active

\defÝ{\. I}

\catcode`ö=\active

\defö{\"{o}}

\catcode`Ö=\active

\defÖ{\"O}

\catcode`ü=\active

\defü{\"{u}}

\catcode`Ü=\active

\defÜ{\"{U}}

\catcode`Þ=\active

\defÞ{\c{S}}

\catcode`þ=\active

\defþ{\c{s}}

\catcode`ý=\active

\defý{{\i}}

\catcode`ç=\active

\defç{\d{c}}

\catcode`Ç=\active

\defÇ{\d{C}}

\begin{document}

\title{Approximate Solutions of Dirac Equation with Hyperbolic-type Potential}
\author{\small Altuð Arda}
\email[E-mail: ]{arda@hacettepe.edu.tr}\affiliation{Department of
Physics Education, Hacettepe University, 06800, Ankara,Turkey}
\author{\small Ramazan Sever}
\email[E-mail: ]{sever@metu.edu.tr}\affiliation{Department of
Physics, Middle East Technical  University, 06531, Ankara,Turkey}

\begin{abstract}
The energy eigenvalues of a Dirac particle for the hyperbolic-type potential field have been computed approximately. It is obtained a transcendental function of energy, $\mathcal{F}(E)$, by writing in terms of confluent Heun functions. The numerical values of energy are then obtained by fixing the zeros on "$E$-axis" for both complex functions $Re[\mathcal{F}(E)]$ and $Im[\mathcal{F}(E)]$.
\end{abstract}


\maketitle

\newpage

\section{Introduction}
The Schrödinger equation is the starting point to taking into account the quantum mechanical effects in physical systems. The Klein-Gordon (KG) and Dirac equations have to be investigated if one also would study the relativistic effects on them. It is well known that the potential forms for which KG and Dirac equations can be solved exactly are very restricted. So, the approximate solutions of the above equations for equal and/or unequal scalar and vector potentials become important to achieve some informations about physical system. The potential fields having exponential shapes are one of the important part having exact or approximate solutions for KG and Dirac equations, and have received great attention in the last few decades [1-15].

The case where scalar and vector potentials are nearly equal in magnitude with opposite sign, $S(r)\simeq -V(r)$, is called the pseudospin symmetry, and the Dirac equation has pseudospin symmetric solutions in that case [16, 17]. If scalar and vector potentials satisfy the condition $S(r)\simeq V(r)$ then the Dirac Hamiltonian has the spin symmetry [18]. The pseudospin symmetry is an exact symmetry for Dirac Hamiltonian under the condition $d(S(r)+V(r))/dr=0$ while the spin symmetry is an exact one under the condition $d(-S(r)+V(r))/dr=0$ [19, 20]. We present here the approximate analytical solutions of Dirac equation for the case where $S(r)\simeq -V(r)$ corresponding to pseudospin symmetry for a potential field which could be written in a exponential form. This hyperbolic-type potential has been studied for the non-relativistic case [21, 22], and the effects of position-dependent mass on spectrum have also been computed [23]. The explicit form of the potential (Fig. 1) is 
\begin{eqnarray}
V(r)=-V_{0}\frac{sinh^{p}(r/2d)}{cosh^{q}(r/2d)}\,,
\end{eqnarray}

\noindent where the potential parameters $V_{0}$ and $d$ correspond to depth, and width of potential, respectively. Becasue of the $p$ and $q$ the above expression defines a class of potentials such as $q=-2, 0, 2, 4, 6$ and $p=-2, 0, \ldots q$. The case $(q, p)=(2, 0)$ describes the Pöschl-Teller potential. For the potentials with $q=-2, 0, 2$, the Schrödinger equation can be converted into the type of hypergeometric equations [21]. The Schrödinger equation for the case where $q=4, 6$ could be written in the form of a confluent Heun equation [24]. In the present work, we achieve also a confluent Heun differential equation [25, 26] by converting the Dirac equation for the case where $(q, p)=(6, 4)$ by using suitable transformations on coordinate variable $r$.

The plan of this paper is as follows. In Section 2, we give the basic equations briefly which also contain some mathematical properties of the solutions of confluent hypergeometric-type equation. In Section 3, we handle a transcendental equation which could be solved numerically for the bound states. This equation is written in terms of the wave functions, and we provide the zeros of the real and imaginary parts of transcendental equation which correspond to the points on "$E$-axis" where the two curves cross. We finally give our conclusions.

\section{The Outline of Dirac Equation}

The Dirac equation for a particle with mass $\mu'$ is written as [27]
\begin{eqnarray}
\left[\vec{\alpha}.\vec{p}+\beta[\mu'+S(r)]+V(r)-E\right]\Psi(r)=0\,,
\end{eqnarray}

\noindent where $E$ is the energy, $\vec{p}$ is the linear momentum, $\vec{\alpha}$ and $\beta$ are $4 \times 4$ matrices have following form, respectively, [27]
\begin{eqnarray}
\vec{\alpha}=\Bigg(\begin{array}{cc}
  0 & \sigma \\
  \sigma & 0
\end{array}\Bigg),\,\,\,\,\,\,\,
\beta=\Bigg(\begin{array}{cc}
  0 & I \\
  -I & 0
\end{array}\Bigg)\,,\nonumber
\end{eqnarray}

\noindent Here, $\sigma$ is Pauli matrix, and $I$ corresponds to $2 \times 2$ unit matrix. The spherically symmetric Dirac spinor is written in terms of the quantum numbers $n$ (usual quantum number), $\kappa$ (spin-orbit quantum number), $\ell$ (orbital angular momentum quantum number), $m$ ($z$-component of angular momentum quantum number), and has the following form
\begin{eqnarray}
\Psi(r)=\,\frac{1}{r}\,\Bigg(\begin{array}{c}
  \,F(r)Y_{jm}^{\ell}(\theta,\phi) \\
iG(r)Y_{jm}^{\tilde{\ell}}(\theta,\phi)
\end{array}\Bigg)\,,
\end{eqnarray}

\noindent where $\tilde{\ell}=\ell+1$, and $\kappa=\pm\,(j+1/2)$ with $j=\ell\pm1/2$ (total angular momentum quantum number). The functions $Y_{jm}^{\ell}(\theta,\phi)$ and $Y_{jm}^{\tilde{\ell}}(\theta,\phi)$ are spherical harmonics, and $F(r)/r$, and $G(r)/r$ are radial part of the Dirac spinor. By inserting Eq. (3) into Eq. (2) we have two coupled differential equations
\begin{subequations}
\begin{align}
\Bigg(\,\frac{d}{dr}\,+\,\frac{\kappa}{r}\,\Bigg)\,F(r)-[\mu'+E-V(r)+S(r)]G(r)=0\,,\\
\Bigg(\,\frac{d}{dr}\,-\,\frac{\kappa}{r}\,\Bigg)\,G(r)-[\mu'-E+V(r)+S(r)]F(r)=0\,,
\end{align}
\end{subequations}

By using Eq. (4a) for $F(r)$, and substituting it into Eq. (4b), we obtain two independent, second-order differential equations as
\begin{subequations}
\begin{align}
\Big\{\,\frac{d^2}{dr^2}\,-\,\frac{\kappa(\kappa-1)}{r^2}+(\mu'-E+C)[V(r)-S(r)]\,\Big\}\,G(r)=[\mu'^2-E^2+C(\mu'+E)]G(r)\,,\\
\Big\{\,\frac{d^2}{dr^2}\,-\,\frac{\kappa(\kappa+1)}{r^2}-(\mu'+E-C')[V(r)+S(r)]\,\Big\}\,F(r)=[\mu'^2-E^2-C'(\mu'-E)]F(r)\,,
\end{align}
\end{subequations}
\noindent where the constants $C=V(r)+S(r)$ (i.e., $dC/dr=0$), and $C'=V(r)-S(r)$ (i.e., $dC'/dr=0$), respectively. In the next section, we deal with the solutions of Eq. (5a) for the potential under consideration.

\section{Hyperbolic-type Potential}

Inserting the potential with $(6, 4)$ into Eq. (5a), taking into account $V(r)+S(r)=C$, and using the approximation instead of the centrifugal term
\begin{eqnarray}
\frac{1}{r^2} \simeq \frac{1}{4sinh^2(r/2d)}\,,\nonumber
\end{eqnarray}

\noindent gives us
\begin{eqnarray}
\left[\frac{d^2}{dx^2}-4d^2\left(\frac{\kappa(\kappa-1)}{4sinh^2(x)}-MV_{0}\,\frac{sinh^4(x)}{cosh^6(x)}-\varepsilon\right)\right]G(x)=0\,,
\end{eqnarray}

\noindent where $x=r/2d$, $M=\mu'-E+C$, and $\varepsilon=\mu'^2-E^2+C(\mu+E)$.

By making a new transformation $y=1/cosh^2(x)$, writing the wave function $G(y)=e^{Ay}f(y)$, and with the help of following abbreviations
\begin{subequations}
\begin{align}
&a_{1}=-\frac{1}{4}\,d^2\kappa(\kappa-1)-d^2MV_{0}-d^2\varepsilon\,,\\
&a_{2}=a_{1}+d^2MV_{0}\,,
\end{align}
\end{subequations}

\noindent we obtain a differential equation for $f(y)$ as
\begin{eqnarray}
\Big\{\frac{d^2}{dy^2}+\left(2A+\frac{1}{y}-\frac{1/2}{1-y}\right)\frac{d}{dy}+\left[\frac{A+a_1}{y}+\frac{a_2-A/2}{1-y}-
\frac{d^2\varepsilon}{y^2}-\frac{d^2\kappa(\kappa-1)/4}{(1-y)^2}\right]\Big\}f(y)=0\,,\nonumber\\
\end{eqnarray}

\noindent with $A^2=-d^2MV_{0}$.

Finally, the transformation such as $f(y)=y^{p'}(1-y)^{q'}H(y)$ gives us a second-order differential equation
\begin{eqnarray}
&&\Big\{\frac{d^2}{dy^2}+\left(2A+\frac{1+2p'}{y}-\frac{2q'+1/2}{1-y}\right)\frac{d}{dy}\nonumber\\
&&+\left[\left(2Ap'+A+a_{1}-2p'q'-q'-\frac{p'}{2}\right)\frac{1}{y}+\left(a_{2}-\frac{A}{2}-2Aq'-2p'q'-q'-\frac{p'}{2}\right)\frac{1}{1-y}
\right]\Big\}H(y)=0\,,\nonumber\\
\end{eqnarray}

\noindent where the constant $p'$ and $q'$ in this equation should satisfy the equalities $p'(p'-1)+p'-d^2\varepsilon=0$, and $q'(q'-1)+q'/2-(1/4)d^2\kappa(\kappa-1)=0$, respectively, for getting a confluent Heun-type equation [25, 26]. If we write Eq. (9) as
\begin{eqnarray}
\Bigg\{\frac{d^2}{dy^2}+\left(2A+\frac{1+B_{1}}{y}+\frac{1+B_{2}}{y-1}\right)\frac{d}{dy}+\frac{B_{3}y+B_{4}}{y(y-1)}\Bigg\}H(y)=0\,,
\end{eqnarray}

\noindent where
\begin{eqnarray}
&B_{1}=2p'\,\,;B_{2}=2q'-(1/2)\,\,;B_{3}=2A(p'+q')+a_{1}-a_{2}+3A/2\nonumber\\&B_{4}=(1+2p')(q'-A)-a_{1}+p'/2\,,
\end{eqnarray}

\noindent and compare with the form [26]
\begin{eqnarray}
\Bigg\{\frac{d^2}{d\rho^2}+\left(\alpha+\frac{\beta+1}{\rho}+\frac{\gamma+1}{\rho-1}\right)\frac{d}{d\rho}
+\left(\frac{\mu}{\rho}+\frac{\nu}{\rho-1}\right)\Bigg\}H(\rho)=0\,,
\end{eqnarray}

\noindent we write the solution of Eq. (10) as $H(y)=HeunC(\alpha,\beta,\gamma,\delta,\eta,y)$. Eq. (10) has two regular singularities at $y=0$, and $1$, and an irregular singularity at $y \rightarrow \infty$. So, the solution is given by [26]
\begin{eqnarray}
H(y)=HeunC(\alpha,\beta,\gamma,\delta,\eta,y)=\sum_{i=1}^{\infty}d_{n}y^{n}\,,
\end{eqnarray}

\noindent where the coefficients $d_{n}$ satisfy the relation $A'_{n}d_{n}=B'_{n}d_{n-1}+C'_{n}d_{n-2}$ with the conditions $d_{0}=1$ and $d_{-1}=1$ with initial conditions $d_{-1}=0$ and $d_{0}=1$. The coefficients in this summation are
\begin{subequations}
\begin{align}
A'_{n}&=1+\beta/n\,,\\
B'_{n}&=1+(\beta+\gamma-\alpha-1)/n+(\eta-\beta)/2+(\gamma-\alpha)(\beta+1)/n^2\,,\\
C'_{n}&=\alpha(\delta/\alpha+(\beta+\gamma)/2+n-1)/n^2\,.
\end{align}
\end{subequations}
where $A'_{n} \rightarrow 1$, $B'_{n} \rightarrow 1$  and $C'_{n} \rightarrow 0$  if $n \rightarrow \infty$. It is assumed that the normalization of $HeunC(\alpha,\beta,\gamma,\delta,\eta,y)=1$. The parameters introduced in last equation satisfy the relations
\begin{subequations}
\begin{align}
\mu&=(1/2)(\alpha-\beta-\gamma+\beta(\alpha-\gamma))-\eta\,,\\
\nu&=(1/2)(\alpha+\beta+\gamma+\gamma(\alpha+\beta))+\delta+\eta\,.
\end{align}
\end{subequations}
In the present work, we have the set of equations about parameters
\begin{eqnarray}
&\alpha=2A\,\,;\beta=B_{1}\,\,;\gamma=B_{2}\,\,;\mu=-B_{4}\,\,;\nu=B_{3}+B_{4}\,,\nonumber\\
&\eta=A(1+2B_{1})+B_{4}-\frac{1}{2}\,(B_{1}+B_{2}+B_{1}B_{2})\,\,;\delta=B_{3}-A(2+B_{1}+B_{2})\,,
\end{eqnarray}

We need two conditions as follows [24-26]
\begin{subequations}
\begin{align}
&\delta=-\alpha\left(N+1+\frac{\beta+\gamma}{2}\right)\,,\\
&\Delta_{N+1}=0\,,
\end{align}
\end{subequations}

\noindent to obtain a confluent Heun polynomial of degree $N \geq 0$ for $y$ (see Ref. [21], for details). $\Delta_{N+1}$ in Eq. (17b) is a $(N+1) \times (N+1)$ determinant. In the light of the above requirements, we write two linearly independent solutions [28]
\begin{subequations}
\begin{align}
&G^{(1)}(x) \sim exp[Asech^2(x)]sech^{2p'}(x)tanh^{2q'}(x)HeunC(\alpha,\beta,\gamma,\delta,\eta,sech^2(x))\,,\\
&G^{(2)}(x) \sim exp[Asech^2(x)]sech^{2(p'-B_{1})}(x)tanh^{2q'}(x)HeunC(\alpha,\beta,\gamma,\delta,\eta,sech^2(x))\,,
\end{align}
\end{subequations}


The confluent Heun functions $HeunC(\alpha,\beta,\gamma,\delta,\eta,y)$ are convergent within $|y|<1$, and asymptotic behaviour at $y=1$ is not known [21, 23-28]. So, by expanding the solution about the other singular point we could obtain an analytical continuation for the Heun function. We find a transcendental function corresponding also the Wronskian written in terms of wave functions [25, 28] which could be solved numerically for relating two solutions. The other set of solutions can be obtained by defining a new variable $z=1-y$, in this instance Eq. (10) becomes
\begin{eqnarray}
\Bigg\{\frac{d^2}{dz^2}+\left(-2A+\frac{1+B_{1}}{z-1}+\frac{1+B_{2}}{z}\right)\frac{d}{dz}
+\left(-\frac{B_{3}+B_{4}}{z}+\frac{B_{4}}{z-1}\right)\Bigg\}H(z)=0\,,
\end{eqnarray}

\noindent with two linearly independent solutions
\begin{subequations}
\begin{align}
&G^{(3)}(x) \sim exp[Asech^2(x)]sech^{2p'}(x)tanh^{2q'}(x)HeunC(-\alpha,\gamma,\beta,-\delta,\eta+\delta,tanh^2(x))\,,\\
&G^{(4)}(x) \sim exp[Asech^2(x)]sech^{2(p'-B_{1})}(x)tanh^{2q'}(x)HeunC(-\alpha,-\gamma,\beta,-\delta,\eta+\delta,tanh^2(x))\,,
\end{align}
\end{subequations}

\noindent Within the quantum mechanics, we should construct mathematical equalities for the wave functions satisfying the continuity conditions which will be discussed in next section.

\section{Bound States}

We could require that two solutions, $G^{(1)}(x)$ and $G^{(3)}(x)$, are sufficient to obtain the energy spectrum. Quantum mechanical restrictions  say us that the wave functions and its derivatives must be continuous such as
\begin{eqnarray}
G^{(1)}(x_)|_{x_{1}}=G^{(3)}(x)|_{x_{1}}\,,\nonumber
\end{eqnarray}
and
\begin{eqnarray}
\frac{dG^{(1)}(x)}{dx}|_{x_{1}}=\frac{dG^{(3)}(x)}{dx}|_{x_{1}}\,,\nonumber
\end{eqnarray}
which could be combined as a determinant
\begin{eqnarray}
\begin{vmatrix} G^{(1)}(x_1) & G^{(3)}(x_1) \\ \frac{dG^{(1)}(x)}{dx}|_{x_{1}} & \frac{dG^{(3)}(x)}{dx}|_{x_{1}} \end{vmatrix}=0\,,
\end{eqnarray}

\noindent where providing a point $(x_1)$ in analytical domains of both $G^{(1)}(x)$ and $G^{(3)}(x)$. Eq. (21) gives us a transcendental equation, and we write for nontrivial solutions
\begin{eqnarray}
\mathcal{F}(E)=G^{(1)}(x)|_{x_{1}}\left(\frac{dG^{(3)}(x)}{dx}\right)|_{x_{1}}-G^{(3)}(x)|_{x_{1}}\left(\frac{dG^{(1)}(x)}{dx}\right)|_{x_{1}}=0\,,
\end{eqnarray}

Since the function $\mathcal{F}(E)$ is complex, the independent equations $Re[\mathcal{F}(E)]=0$ and $Im[\mathcal{F}(E)]=0$ should be solved to obtain real solutions (bound states energies). These two equations can be solved numerically because they are transcendental equations and the energy eigenvalues are the points which correspond to the one on "$E$-axis" where plots of $Re[\mathcal{F}(E)]=0$ and $Im[\mathcal{F}(E)]=0$ are cross. In Fig. (2), we show the dependence of $\mathcal{F}(E)$ on $E$ with $x_1=1$ for the potential parameters $V_{0}=200$, and $d=0.5$. The trigonometric potential has three states at points $E=-6.60, -11.83$ and $-16.13$ for $\mu'=2.0$, while the ones at points $E=-5.99, -11.26$ and $-15.63$ for $\mu'=0.5$, and also at points $E=-5.88, -11.12$ and $-15.53$ for $\mu'=0.01$, respectively. The values of bound state energies shift to left on "$E$-axis" while the mass increases which means that the energy levels appear in more deeper of potential-well.

\section{Conclusion}

The hyperbolic-type potential has been solved for a Dirac particle under the case where scalar and vector potentials are equal in magnitude with opposite in sign. The solutions have been computed in terms of confluent Heun functions. The bound state energies have been found numerically by obtaining the zeros of transcendental function $\mathcal{F}(E)$ on "$E$-axis". It has been given two plots of $\mathcal{F}(E)$ for different mass values to see the bound state energies clearly.

\section{Acknowledgments}
This research was partially supported by the Scientific and
Technical Research Council of Turkey. One of authors (A.A.) thanks Prof Andreas Fring from City University London and the Department of Mathematics for hospitality where the last part of this work has been done.

\newpage

\begin{figure}
\centering
\includegraphics[height=4in, width=6.5in, angle=0]{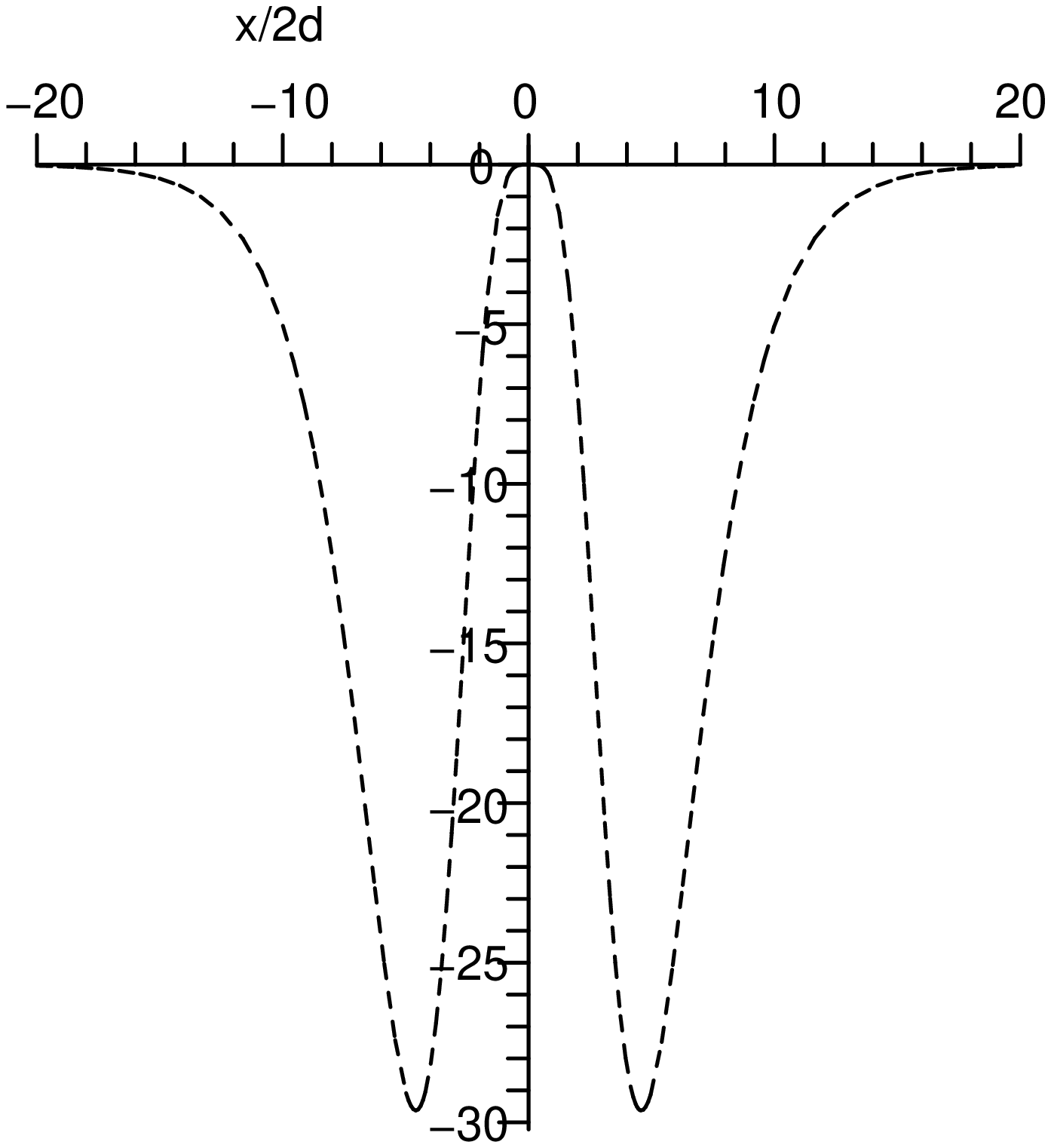}
\caption{The hyperbolic-type potential for $V_{0}=200, d=0.5$.}
\end{figure}

\newpage 

\begin{figure}
\centering \subfloat[][Intersections of transcendental functions
for $\mu'=2$.]{\includegraphics[height=3in,
width=3in, angle=0]{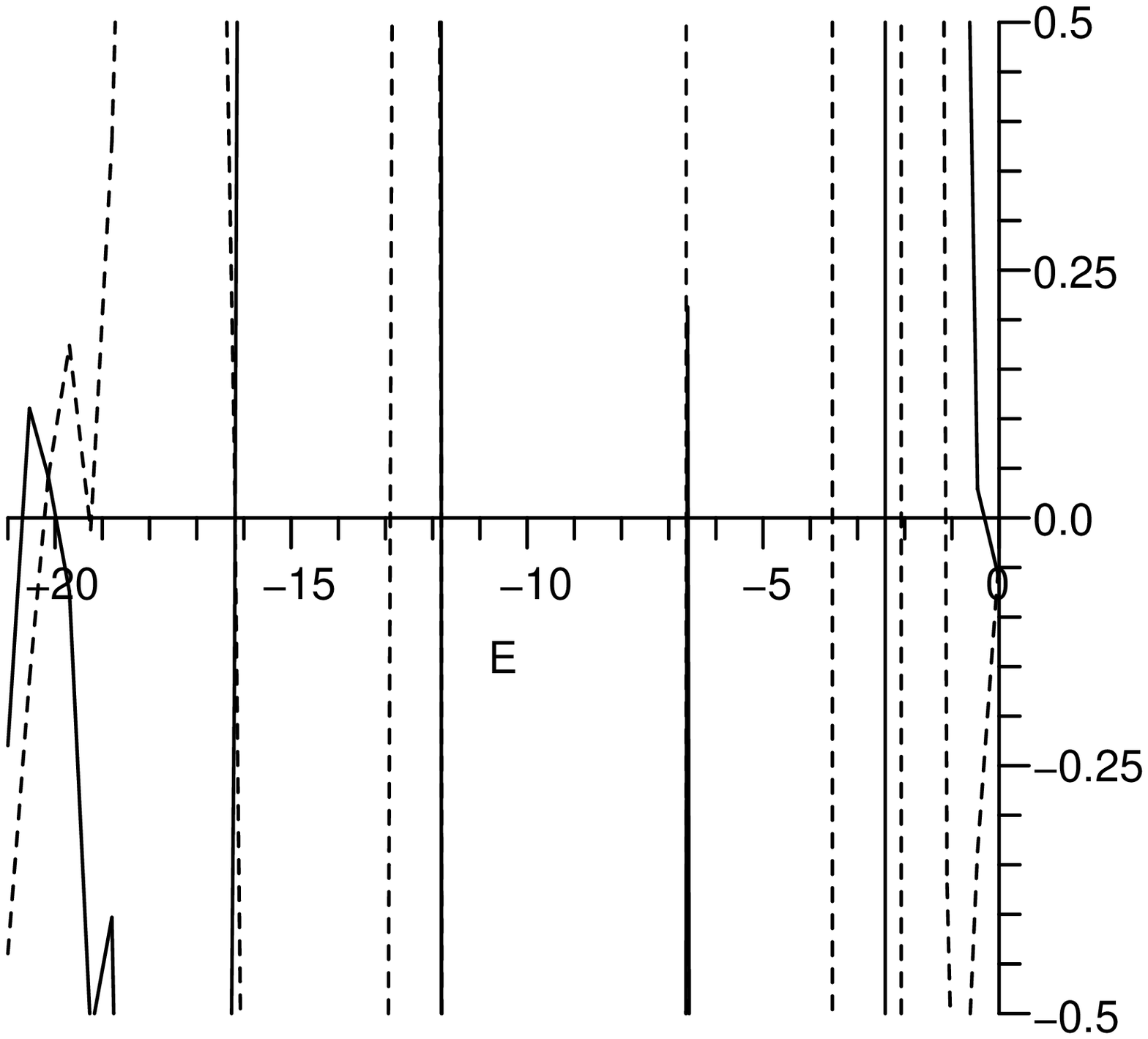}}
\subfloat[][Intersections of transcendental functions
for $\mu'=0.5$.]{\includegraphics[height=3in, width=3in,
angle=0]{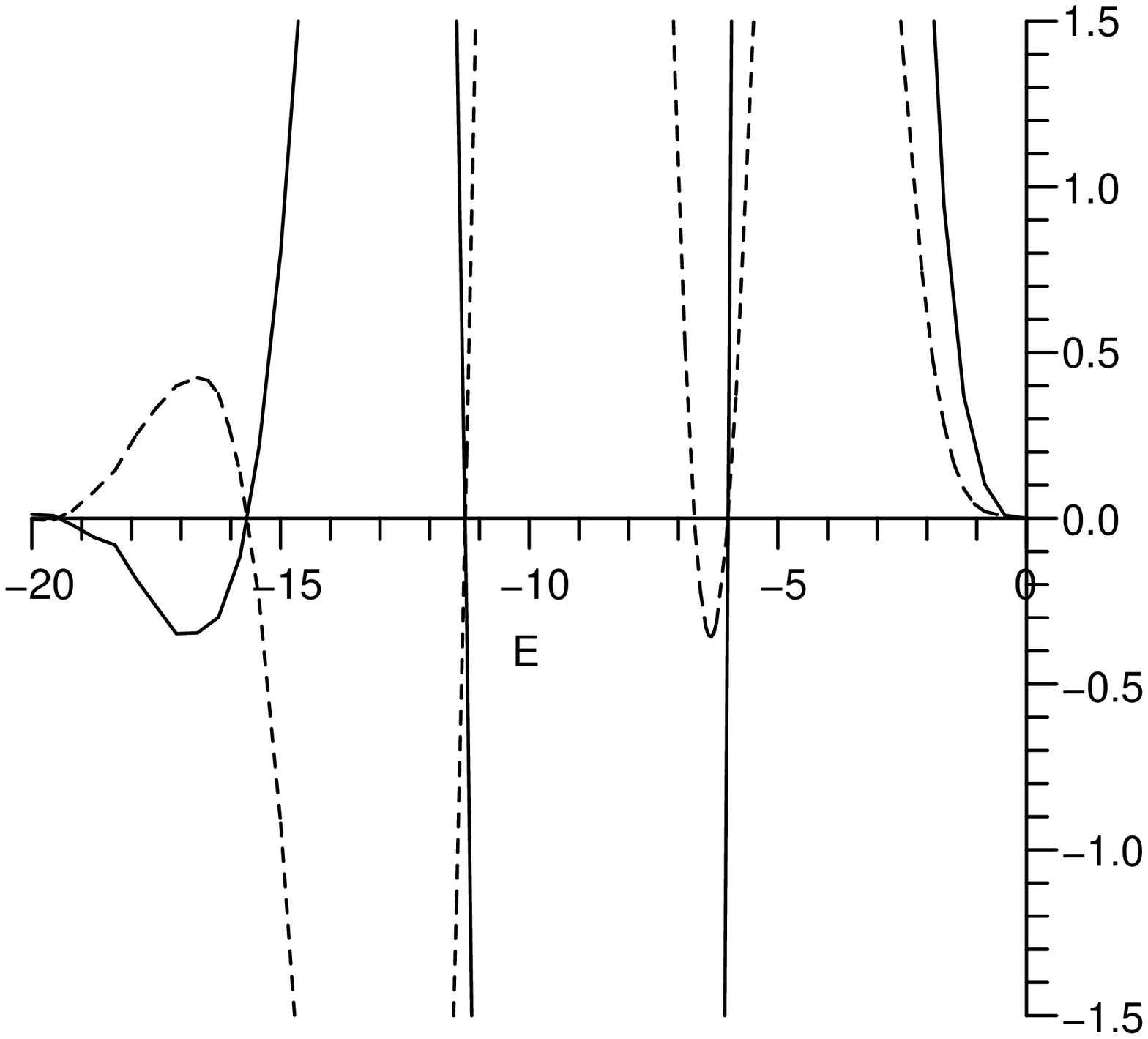}}\\
 \subfloat[][Intersections of transcendental functions
for $\mu'=0.01$.]{\includegraphics[height=3in,
width=3in, angle=0]{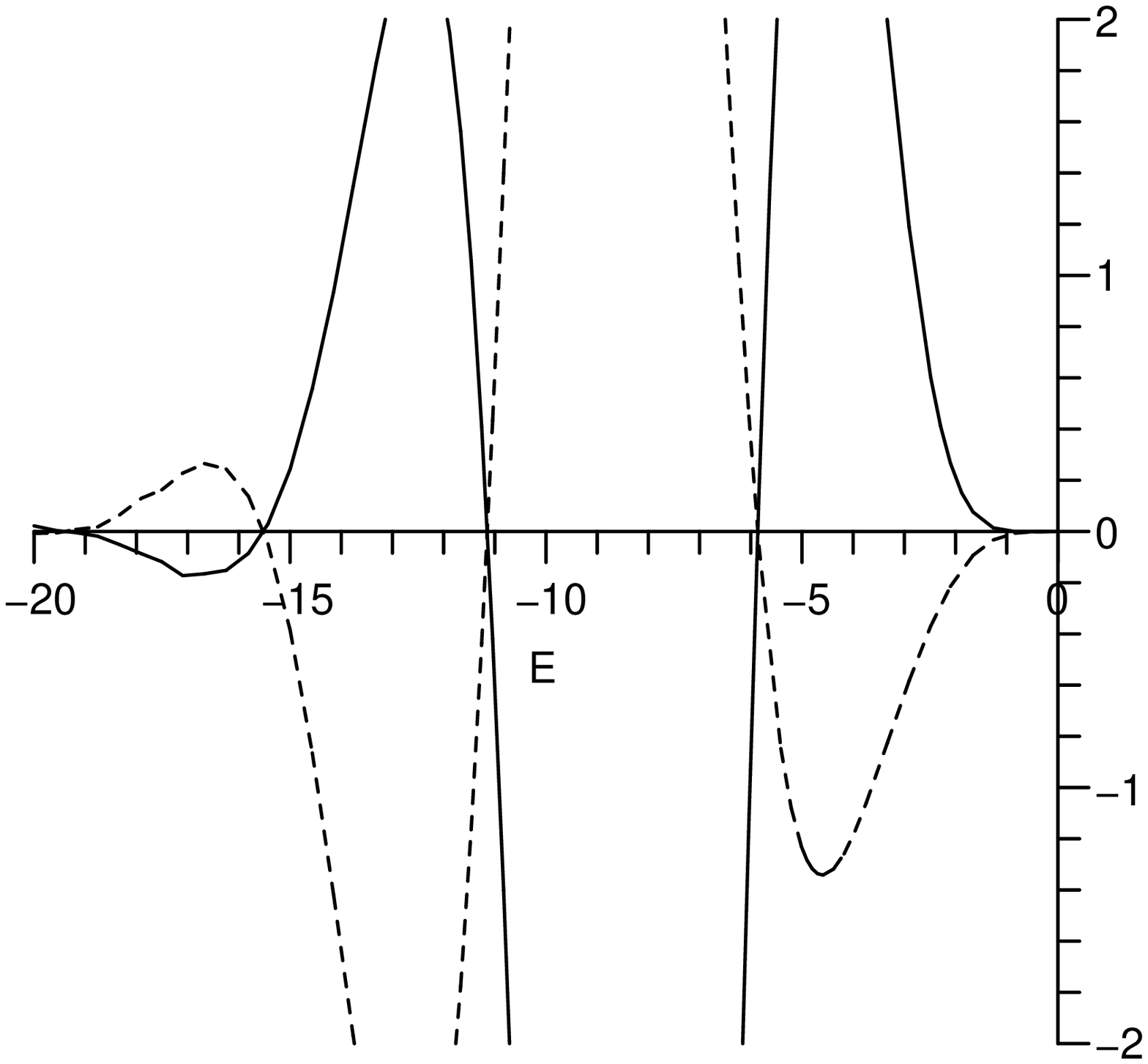}}
 \caption{The energy spectrum is derived by solving numerically the transcendental equations $Re[\mathcal{F}(E)]=0$ and $Im[\mathcal{F}(E)]=0$. In the figure, we show $Re[\mathcal{F}(E)]$ (solid line) and $Im[\mathcal{F}(E)]$ (dashed line). The energy eigenvalues are given by the points on horizontal line where two curves cross. Parameters: $\kappa=1, C=0, \hbar=c=1$.}
\end{figure}

\end{document}